\documentclass[apj]{emulateapj}
\usepackage{graphicx}
\usepackage{natbib}
\usepackage{color}      
\usepackage{appendix}

\shorttitle{\emph{HST} STIS UV Spectroscopy of Hen\,3-1475}
\shortauthors{Fang et al.}

\begin{document}

\title{\emph{HST} STIS UV Spectroscopic Observations of the 
Protoplanetary Nebula Hen\,3-1475\footnotemark[$\ast,**$]}

\footnotetext[$\ast$]{Based on observations made with the NASA/ESA 
\emph{Hubble Space Telescope}, obtained at the Space Telescope 
Science Institute, which is operated by the Association of 
Universities for Research in Astronomy, Inc., under NASA contract 
NAS\,5-26555.  The UV observations are associated with the 
\emph{HST} GO program \#13838.} 

\footnotetext[$**$]{Dedicated to Dr.\ Angels Riera, who passed 
away on 2017 September 27 in Barcelona, Spain.  Dr.\ Riera was 
the first to recognize the peculiar nature of Hen\,3-1475.} 

\author{Xuan Fang$^{1,2}$\footnotemark[$\dagger$], 
Ana I.\ G\'{o}mez~de~Castro$^{3}$, 
Jes\'{u}s A.\ Toal\'{a}$^{4}$, 
and 
Angels Riera$^{5}$\footnotemark[$\ddag$]} 
\affil{
$^{1}$Laboratory for Space Research, Faculty of Science, The 
University of Hong Kong, Pokfulam Road, Hong Kong, P.~R.\ China\\
$^{2}$Department of Physics, Faculty of Science, The University of 
Hong Kong, Pokfulam Road, Hong Kong, P.~R.\ China\\
$^{3}$AEGORA Research Group, Facultad de Ciencias, Universidad 
Complutense, E-28040, Madrid, Spain\\
$^{4}$Instituto de Radioastronom\'{i}a y Astrof\'{i}sica, UNAM 
Campus Morelia, Apartado postal 3-72, Morelia 58090, Michoac\'{a}n,
Mexico\\
$^{5}$Departament de F\'\i sica i Enginyeria Nuclear, EUETIB, 
Universitat Polit\'{e}cnica de Catalunya, 
E-08036 Barcelona, Spain} 

\footnotetext[$\dagger$]{Visiting Astronomer, Key Laboratory of 
Optical Astronomy, National Astronomical Observatories, Chinese 
Academy of Sciences (NAOC), 20A Datun Road, Chaoyang District, 
Beijing 100101, P.~R.\ China}

\footnotetext[$\ddag$]{Deceased.}

\email{fangx@hku.hk}

\begin{abstract} 
We present UV spectra of the protoplanetary nebula (pPN) 
Hen\,3-1475 obtained with the Space Telescope Imaging Spectrograph 
(STIS) on board the \emph{Hubble Space Telescope} (\emph{HST}). 
Our deep, low-dispersion spectroscopy enables monochromatic imaging
of Hen\,3-1475 in UV nebular emission lines, the first of such 
attempt ever made for a pPN.  The high spatial resolution of STIS 
imaging allows an unprecedentedly sharp view of the $S$-shaped 
jet, especially the inner NW1 knot, which is resolved into four 
components in Mg\,{\sc ii} $\lambda$2800.  Through critical 
comparison with \emph{HST} optical narrowband images, we found a 
negative radial velocity gradient in NW1, from $-$1550~km\,s$^{-1}$
on its innermost component to $\sim-$300~km\,s$^{-1}$ on the 
outermost.  Despite their high radial velocities, these components 
of NW1 mostly show no obvious (or very small) proper motions, 
indicating that they might be quasi-stationary shocks near the tip 
of the conical flow along the collimated jet of Hen\,3-1475. 
\end{abstract}


\section{Introduction}
\label{sec1}

Planetary nebulae (PNe) exhibit a great variety of morphologies. 
The low occurrence of spherical symmetry is perplexing because most
AGB envelopes are spherical \citep[e.g.,][]{neri98,mauron13}.  The
transformation from a spherical AGB envelope to an aspherical PN 
begins at the protoplanetary nebula (pPN) phase, when the stellar 
temperature is still too low to photoionize the nebula.  Collimated
outflows as fast as 400~km~s$^{-1}$ have been detected in pPNe 
\citep{bujarrabal01}, and play a major role in PN shaping 
\citep{st98}.  An even faster jet (630~km\,s$^{-1}$) was found in 
the young PN MyCn\,18 \citep{miszalski18}.  Binarity is believed 
to be responsible for jet formation, as recently observed in the 
symbiotic star R\,Aqr \citep{schmid17} and the AGB star Y\,Gem 
\citep{sahai18}. 

\begin{figure}
\begin{center}
\includegraphics[width=1.0\columnwidth,angle=0]{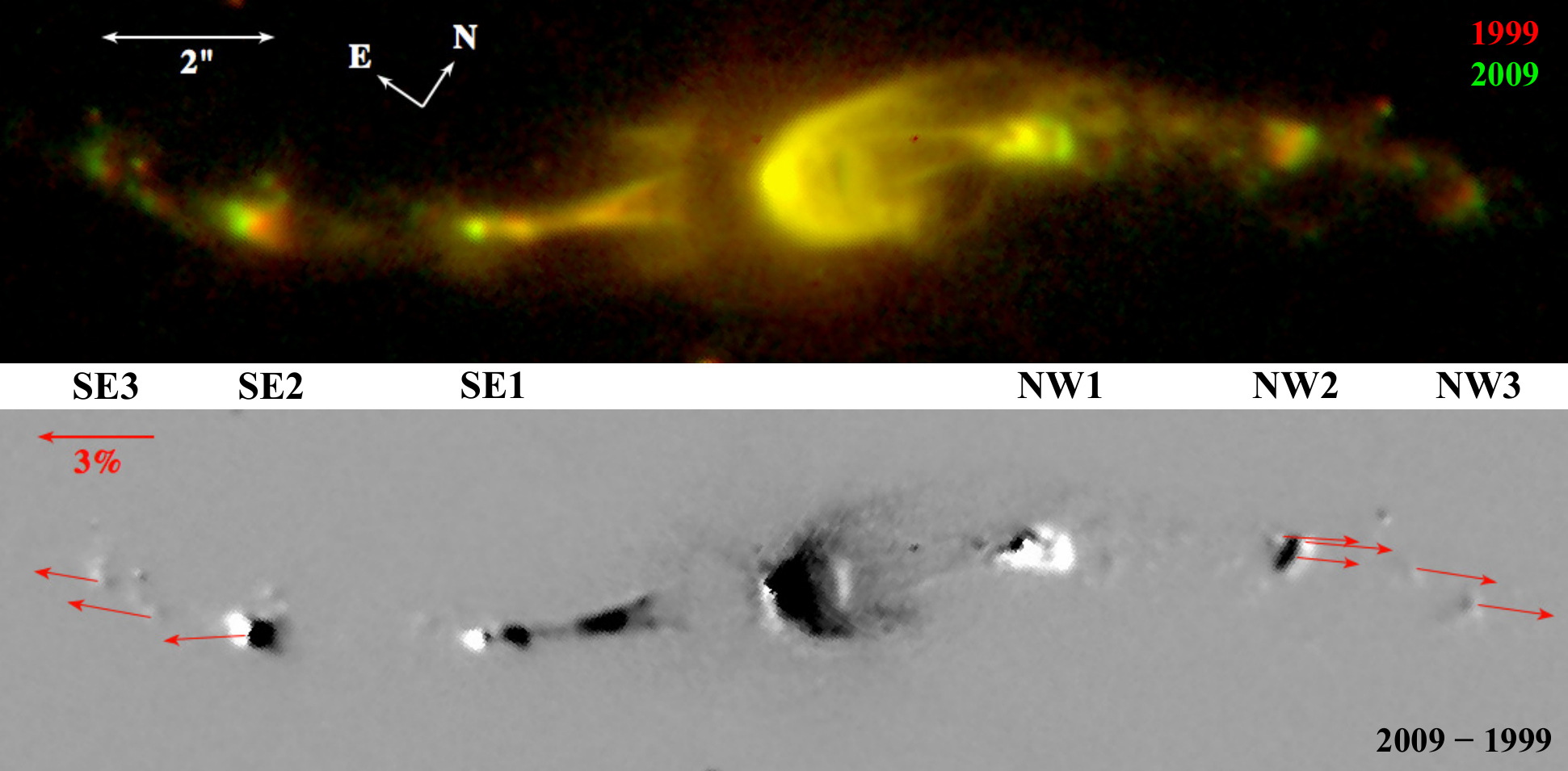}
\caption{Top: \emph{HST} F658N images of Hen\,3-1475 obtained in 
1999 (red) and 2009 (green).  Bottom: residual F658N image (a 
2009$-$1999 difference); red arrows overlaid show the expansion 
directions, with lengths proportional to fractional expansion (in 
percentage, as defined by equation~2 in \citealt{fang14}).} 
\label{fig1}
\end{center}
\end{figure}

Hen\,3-1475 (a.k.a.\ the ``Garden Sprinkler Nebula''), was first 
recognized as a transition object in the post-AGB phase \citep{pp89}
and later found to have fast bipolar outflows \citep{riera95}. 
Since its nature as a pPN was settled, Hen\,3-1475 has been imaged 
on several occasions \citep[][]{bobrowsky95,borkowski97,ueta00,bh01}.
It has a highly collimated bipolar structure with an $S$-shaped 
string of point-symmetric, [N\,{\sc ii}]-bright knots extending 
over $\sim$17\arcsec\ along the main axis (Figure~\ref{fig1}).
High-dispersion spectroscopy revealed high-velocity jets 
\citep[1200~km\,s$^{-1}$,][]{bh01} and ultra-fast (up to 
2300~km\,s$^{-1}$) winds very close to the central star 
\citep{sanchez01}.  To date, Hen\,3-1475 is the only pPN with 
detected diffuse X-ray emission, a signature of stellar winds 
interactions \citep{sahai03}; this X-ray emission comes mostly from 
its brightest NW1\footnote{Following \citet[][]{borkowski97}, we 
hereafter refer the inner, middle, and outer pairs of knots in 
Hen\,3-1475 as NW1/SE1, NW2/SE2, and NW3/SE3, respectively.  See 
also Figure~\ref{fig1}.} knot, whose emission is shock excited 
\citep{bh01,riera03,riera06}.  Its outflows are being collimated 
into jets far away from the central star \citep{borkowski97}, but 
the collimation mechanism is not well understood.  A time-dependent 
ejection velocity model with a period of 1500~yr was proposed to 
explain its morphology and kinematics \citep{velazquez04}. 

\emph{IUE} detection of UV emission lines in Hen\,3-1475 was only 
marginal due to dust obscuration \citep{gp03}; but the effects of 
shocks can be studied in the UV.  Here we report on UV spectroscopy
of Hen\,3-1475 obtained with the \emph{Hubble Space Telescope} 
(\emph{HST}).  After briefly revisiting jet propagation in 
Section~\ref{sec2}, we introduce our \emph{HST} observations in 
Section~\ref{sec3}.  Data analysis and discussion are presented in 
Section~\ref{sec4}, and summary and future plan in 
Section~\ref{sec5}. 

\section{Jet Propagation of Hen\,3-1475}
\label{sec2}

Jet propagation of Hen\,3-1475 was studied by \citet{bh01} based 
on 1996/1999 observations.  We now investigate it using the F658N 
images obtained on 1999 September 9 (WFPC2/PC; prop.~7285, PI: 
J.~P.\ Harrington) and 2009 August 6 (WFC3; prop.~11580, PI: B.\ 
Balick).  The WFPC2/PC (0\farcs050\,pixel$^{-1}$) image was 
aligned, rebinned and scaled to WFC3 (0\farcs0396\,pixel$^{-1}$) 
following the procedure described in \citet{fang14}.  The two 
images displayed together show obvious proper motion of the outer 
knots (NW2/SE2 and NW3/SE3), but it is {\it not obvious} in NW1/SE1 
whose outer parts are much brighter in 2009 than in 1999 
(Figure~\ref{fig1}~top).  This brightening may cause apparently 
large (but false) proper motion of NW1/SE1 
(Figure~\ref{fig1}~bottom). 
Proper motions of the outer knots are $\sim$10--15~mas\,yr$^{-1}$, 
corresponding to sky-projected velocities of 230--360~km\,s$^{-1}$ 
(at a distance of 5\,kpc, \citealt{riera95}; hereafter adopted); 
these are consistent with the measurements of \citet{bh01}.  The 
outer knots move almost \emph{tangentially} along the $S$-shaped 
arms of Hen\,3-1475 (Figure~\ref{fig1}~bottom), unlike the purely 
radial (from the core) motions usually expected for the knots 
ejected from a precessing nozzle as proposed by \citet{riera03}. 
The possible mechanism of such tangential motion is discussed in 
Section~\ref{sec4:2}.  Figure~\ref{fig1} also shows very small 
motion of the two inner cones \citep[i.e., the limb-brightened 
edges of conical shocks;][]{borkowski97} that originate from the 
central torus and ``converge'' at NW1/SE1. 

\begin{figure}
\begin{center}
\includegraphics[width=1.0\columnwidth,angle=0]{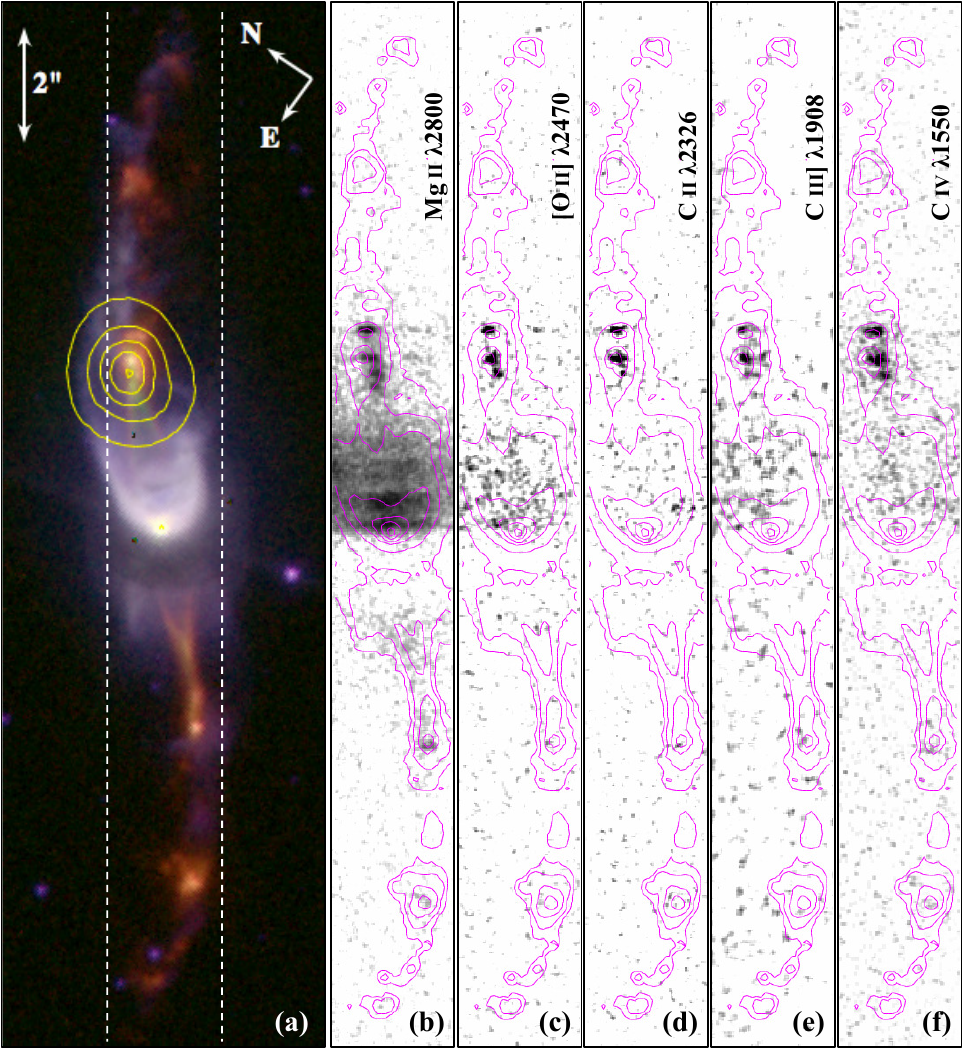}
\caption{Panel~{\bf a}: \emph{HST} 2009 WFC3 color-composite image 
of Hen\,3-1475 created with F658N (red), F656N (green), and F555W 
(blue) and overlaid with \emph{Chandra} ACIS X-ray emission contours
(yellow; Obs.~ID 2580; PI: R.\ Sahai).  White dashed lines indicate 
the STIS 52\arcsec$\times$2\arcsec\ long slit (PA=124\fdg65). 
Panels~{\bf b}--{\bf f}: STIS UV spectral-line images (see legend) 
overlaid by WFC3 F658N emission contours (magenta); each panel is 
2\arcsec$\times$17\farcs2 in size, displayed in negative greyscale 
and slightly smoothed to reduce noise.} 
\label{fig2}
\end{center}
\end{figure}

\section{STIS Observations}
\label{sec3}

Long-slit UV spectra of Hen\,3-1475 were obtained with the Space 
Telescope Imaging Spectrograph (STIS) on board the \emph{HST} on 
2015 June 11, under the GO program \#13838 (PI: X.\ Fang, 
Cycle~22).  First-order gratings G140L and G230L were used to 
obtain the FUV and NUV spectra, respectively.  The STIS 
52\arcsec$\times$2\arcsec\ long slit was placed on the central 
core (R.A.=$17^{\rm h}45^{\rm m}14\rlap{.}^{\rm s}19$, 
Decl.=$-17^\circ56^\prime46\rlap{.}^{\prime\prime}9$) 
of Hen\,3-1475 with a position angle (PA) of 124\fdg65 along the 
nebular axis.  All knots lie within the slit (Figure~\ref{fig2}a). 
The observations were in the TIME-TAG mode that allows us to 
scrutinize the data for periods of high airglow level.  The 
instrument parameters and exposures are all summarized in 
Table\,\ref{obs_log}, where we also present the actual angular 
resolutions (in arcsec) for the two grating modes as provided by 
the STIS Instrument Handbook. 

The data were reduced and calibrated with the \emph{HST} STIS 
pipeline.  The long-slit low-dispersion spectroscopy enables 
\emph{monochromatic imaging} of Hen\,3-1475 on UV nebular emission 
lines (see description in Section~\ref{sec4:1}), whilst 
reducing the impact of the bright geocoronal Ly$\alpha$ 
emission.  Several emission lines of the fully calibrated spectra 
are demonstrated in Figure~\ref{fig2}.  The FUV spectrum includes 
N\,{\sc v} $\lambda\lambda$1239,1242 
(all wavelengths in {\AA}), C\,{\sc iv} $\lambda$1550 (a 
blend of $\lambda\lambda$1548,1551), He\,{\sc ii} $\lambda$1640, 
and O\,{\sc iii}] $\lambda\lambda$1661,1666.  In NUV we detected 
C\,{\sc iii}] $\lambda$1908, C\,{\sc ii} $\lambda$2326, [O\,{\sc 
ii}] $\lambda$2470 and Mg\,{\sc ii} $\lambda$2800 (a blend of 
$\lambda\lambda$2795,2803). 
No fine-structure transition lines in any doublet detected in our 
low-dispersion spectra are spectroscopically resolved.

\begin{table*}
\begin{center}
\caption{\emph{HST} STIS Observing Log}
\label{obs_log}
\begin{tabular}{lcccccccc}
\hline
\hline
Detector & Grating & $\lambda_{\rm c}$ & Spectral Range & Spectral Res. & Dispersion & Spatial Scale & Angular & $t_{\rm exp}$\\
 &  & ({\AA}) & ({\AA}) & ($R$) & ({\AA}~pixel$^{-1}$) & (pixel$^{-1}$) & Resolution & (s)\\
\hline
STIS/FUV-MAMA & G140L & 1425 & 1150--1730 & 960--1440 & 0.58 & 0\farcs0246 & $\sim$0\farcs07 & 2800\\
STIS/NUV-MAMA & G230L & 2376 & 1570--3180 & 500--1010 & 1.55 & 0\farcs0248 & $\sim$0\farcs06 & 2100\\
\hline
\multicolumn{9}{l}{NOTE. -- Instrument parameters are adopted from 
http://www.stsci.edu/hst/stis/design/gratings.}
\end{tabular}
\end{center}
\end{table*}

\section{Results and Discussion}
\label{sec4} 

\subsection{UV Morphology}
\label{sec4:1}

\begin{figure}
\begin{center}
\includegraphics[width=1.0\columnwidth,angle=0]{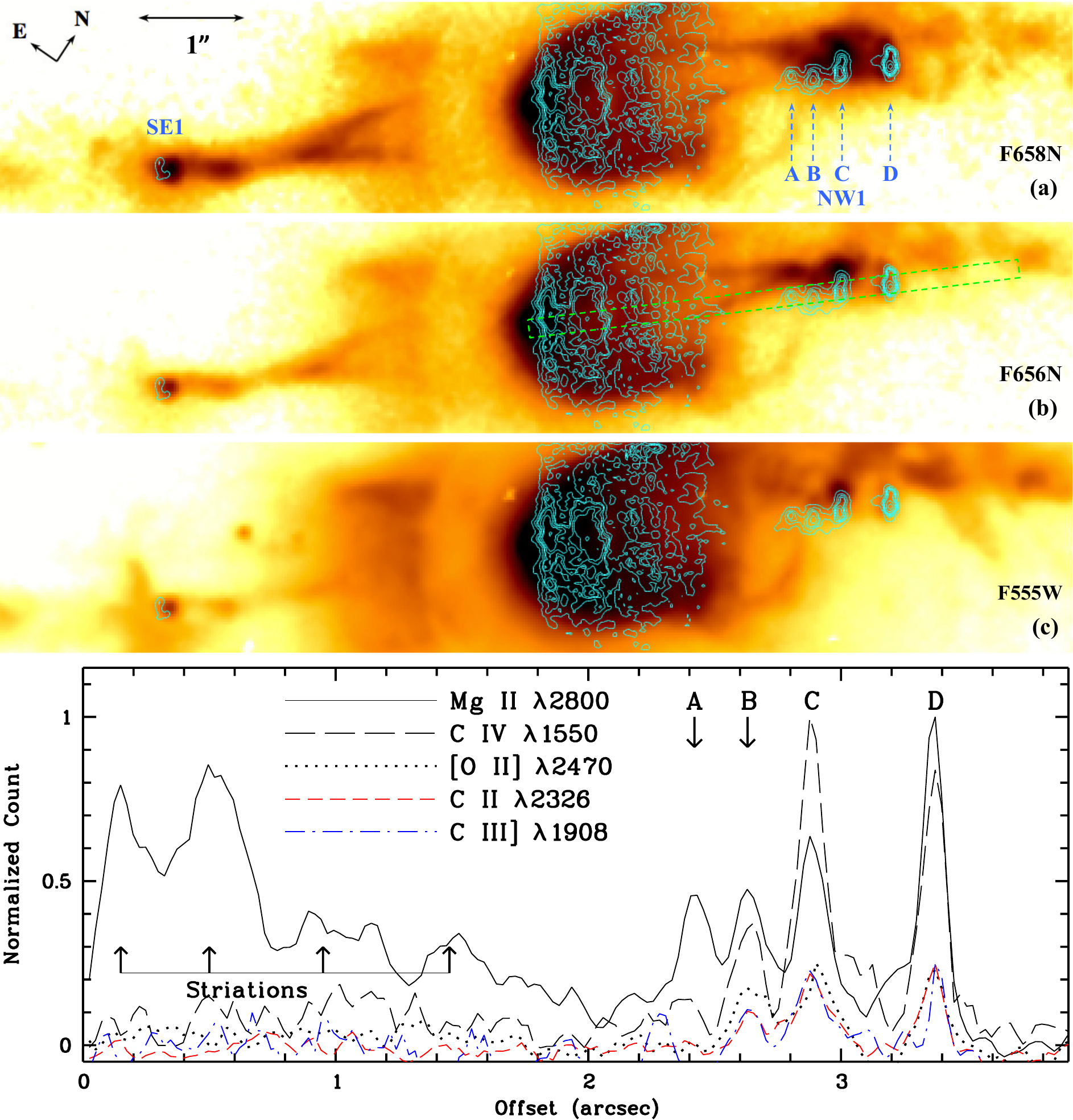}
\caption{Emission contours (cyan) of the STIS Mg\,{\sc ii} 
$\lambda$2800 line overlaid on the WFC3 F658N ({\bf a}), F656N 
({\bf b}), and F555W ({\bf c}) images showing the central 
2\arcsec$\times$10\arcsec\ region of Hen\,3-1475.  The NW1 knot is 
resolved into four components (A, B, C, and D labeled in blue) in 
Mg\,{\sc ii}. 
Nebular core is saturated in all three filters to enhance lobe 
structures, including NW1/SE1 and the conical shocks.  Bottom: 
Emission profiles of different UV lines along a cut (green-dashed 
rectangle in panel {\bf b}) through the central star and the NW1 
knot components, at PA=311\degr; positions of the four components 
of NW1 and the striations seen in Mg\,{\sc ii} emission 
(Figure~\ref{fig2}b) are labeled. Emissions of [O\,{\sc ii}], 
C\,{\sc ii} and C\,{\sc iii}] are normalized and further scaled for 
easy comparison.} 
\label{fig3}
\end{center}
\end{figure}

The UV morphology of Hen\,3-1475 is studied in conjunction with 
\emph{HST} WFC3 images in the F656N and F658N filters that 
show the H$\alpha$ and [N\,{\sc ii}] $\lambda$6583 emission, 
mostly from the jets and knots, which are presumably shock excited 
\citep{bh01}; the broadband images (e.g., F555W and F814W) show 
the continuum radiation, mostly scattered starlight. 
We created UV monochromatic images by first selecting in the STIS 
2D spectrum regions centered on UV emission lines, and then 
carefully registered them with WFC3 images.  The overall morphology
of Hen\,3-1475's brightest regions (NW1/SE1 and the central 
reflection nebula), as outlined in optical emission, is visible in 
several UV lines (Figure~\ref{fig2}). 
In the Mg\,{\sc ii} $\lambda$2800 spectral-line image, the NW1 knot
is resolved into four components, hereafter named A, B, C, and D 
from the inside out (Figure~\ref{fig3}a,\,b), with angular sizes 
of $\sim$0\farcs12--0\farcs14 along the jet axis; in the other UV 
lines, generally fainter than Mg\,{\sc ii}, only the three outer 
components B, C, and D are seen (Figure~\ref{fig2}b--f).  The SE1 
knot seen in the UV could be the counterpart of the outermost D 
component. 

\begin{figure}
\begin{center}
\includegraphics[width=1.0\columnwidth,angle=0]{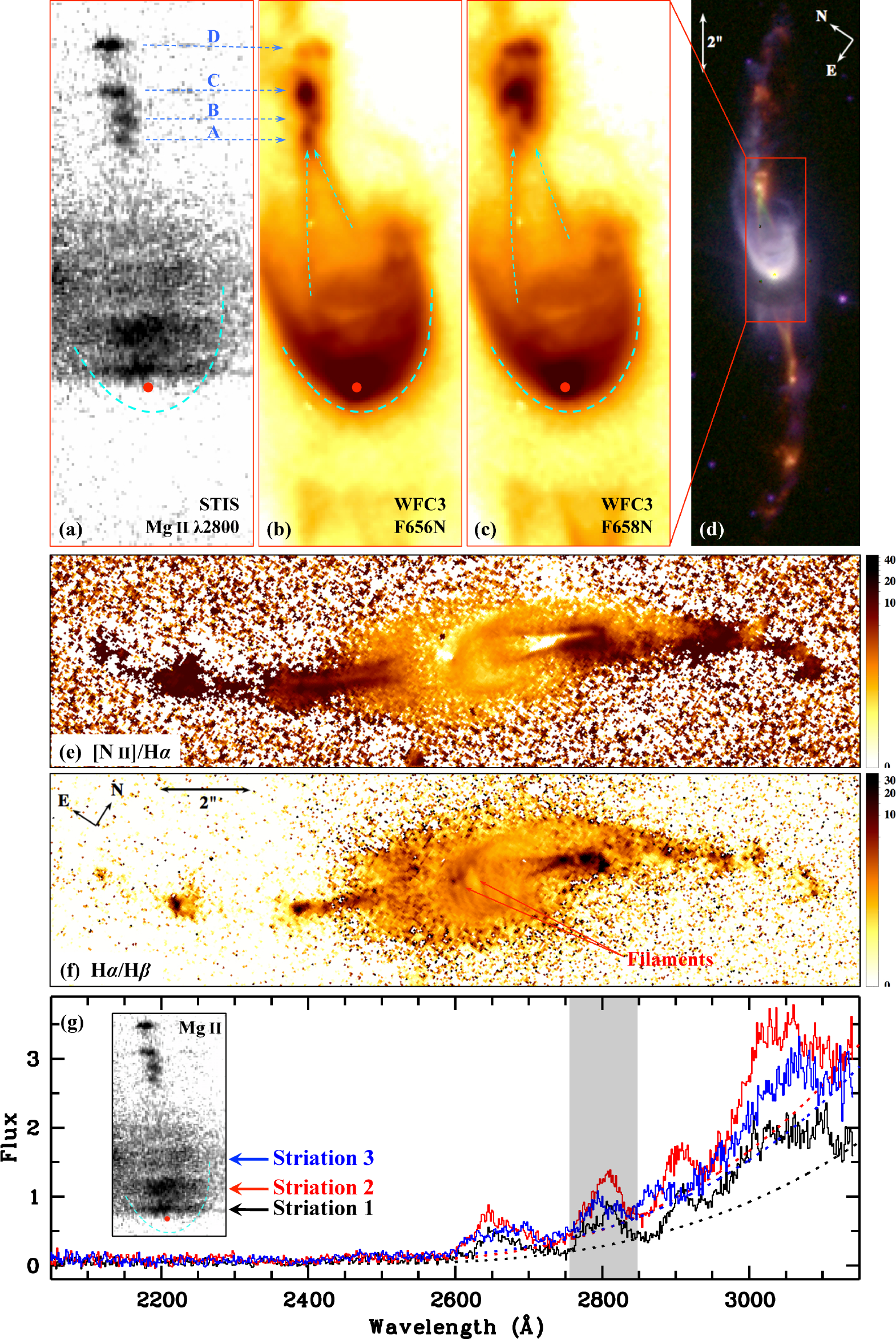}
\caption{Close-up view of the central 2\arcsec$\times$5\farcs4  
region of Hen\,3-1475 in Mg\,{\sc ii} $\lambda$2800 ({\bf a}), 
H$\alpha$ ({\bf b}), and [N\,{\sc ii}] ({\bf c}); the four 
components of the NW1 knot are labeled.  The slightly curved 
conical converging shock and the outer boundary of NW lobe are 
delineated ({\bf b}, {\bf c}).  Red dot marks the position of 
the central star.  Panel {\bf d} is the same as Figure~\ref{fig2}a.
Panels {\bf e} and {\bf f} are [N\,{\sc ii}]/H$\alpha$ and 
H$\alpha$/H$\beta$ ratio images, respectively, both displayed in 
inverted false color (i.e., dark regions in panel {\bf e} have 
strong [N\,{\sc ii}] emission relative to H$\alpha$) and 
accompanied by color bars in logarithmic scale.  Panel {\bf g}: 
Extracted 1D NUV spectra of the three inner striations as indicated
by color-coded arrows on an inset of the Mg\,{\sc ii} $\lambda$2800
spectral-line image (same as panel~{\bf a}); the dotted curves are 
polynomial fits of the corresponding underlying continua; the 
Mg\,{\sc ii} emission feature is marked by the grey-shaded area. 
Flux is in units of 10$^{-14}$ 
erg\,cm$^{-2}$\,s$^{-1}$\,{\AA}$^{-1}$.} 
\label{fig4}
\end{center}
\end{figure}

The optical counterparts (in H$\alpha$) of the four UV components 
A, B, C, and D are located 2\farcs49 (1\farcs73-W, 1\farcs79-N), 
2\farcs67 (1\farcs93-W, 1\farcs85-N), 2\farcs91 (2\farcs07-W, 
2\farcs04-N), and 3\farcs30 (2\farcs47-W, 2\farcs19-N) from the 
central star, respectively.  B, C and D are elongated (in the 
dispersion direction) in Mg\,{\sc ii}; this elongation is obvious 
in B and C, which are compact in H$\alpha$.  D comprises two 
subcomponents in H$\alpha$ that are $\sim$0\farcs15 apart and 
stretch by 0\farcs32, similar to the size in the UV. The NW conical
shock ``converges'' on A (Figure~\ref{fig4}a,\,b), the  most 
compact (radius$\sim$0\farcs06) among the four UV components of 
NW1.  Elongation of the NW1 components in the UV may be caused by 
large, internal negative radial velocity gradients \citep{bh01}. 

Along the (spectral) dispersion direction, the four components of 
NW1 seem to be displaced with respect to their optical counterparts;
we discuss this in Section~\ref{sec4:2}. 
Along the slit direction, positions of A, B, and C in the UV are 
unchanged compared to their counterparts in H$\alpha$, which is 
consistent with the very little expansion of NW1 we found through 
comparison of the 1999/2009 F658N images (see Section~\ref{sec2}). 
Component D and SE1 in the UV are slightly outside their H$\alpha$ 
counterparts (Figure~\ref{fig3}b) due to jet expansion/propagation 
since 2009.  The position of D in Mg\,{\sc ii} is shifted outward 
(i.e., towards NW) by 0\farcs069 from its position in H$\alpha$, 
which corresponds to a proper motion of 11.8~mas\,yr$^{-1}$, or 
$\sim$280~km\,s$^{-1}$. 
Similarly, SE1 in Mg\,{\sc ii} is displaced from its position in 
H$\alpha$ by 0\farcs091, corresponding to $\sim$15.5~mas\,yr$^{-1}$, 
or $\sim$360~km\,s$^{-1}$. 


In the UV spectrum, $<$2\arcsec\ from the central star, there are 
four ``stripes'' (or striations) that are more prominent near 
the red end of the STIS NUV-MAMA spectrum (Figure~\ref{fig2}b). 
These striations are only detected in the NW lobe, and seem to 
spatially align with the optical features at corresponding 
locations: the hollow $U$-shaped ``bowls'' near the nucleus of 
Hen\,3-1475 also seem striated in the optical 
(Figure~\ref{fig4}b,\,c).  The central region of Hen\,3-1475 is 
dominated by the scattered stellar light as shown in the \emph{HST}
STIS G750M spectrum \citep[][Figure~2 therein]{bh01}. 
Along the jet axis, the four UV striations in the NW lobe are 
$\sim$0\farcs13, 0\farcs5, 0\farcs95 and 1\farcs43 from the central
star (Figure~\ref{fig3}~bottom). 

\subsection{Jet Kinematics and Shocks}
\label{sec4:2} 

Whereas the positions of components A, B, and C in UV emission 
along the slit direction are consistent with those of their 
counterparts in H$\alpha$, there is a noticeable shift along the 
dispersion direction of the UV spectrum (Figure~\ref{fig3}b). 
Given that the NW jet of Hen\,3-1475 has high approaching speeds 
\citep{bh01,riera03}, it is plausible to attribute these 
displacements to Doppler shift.\footnote{In [N\,{\sc ii}] emission 
there are NW1 components that seem to spatially coincide better, 
along the SW edges, with the corresponding Mg\,{\sc ii} ones than 
H$\alpha$ does; this indicates that the offsets of these UV 
components with respect to H$\alpha$ might be, at least partially, 
due to Mg\,{\sc ii} emission arising from the [N\,{\sc 
ii}]-emitting gas toward the SW from the H$\alpha$ emission.  An 
in-depth analysis of the UV$+$optical data is needed to confirm 
this.}  Indeed, A, B, and C are shifted blueward in the UV spectrum 
by 9.4, 7.2, and 6.7 pixels which, if entirely due to Doppler shift, 
correspond to radial velocities of $-$1550$\pm$160, $-$1200$\pm$330 
and $-$1100$\pm$400~km\,s$^{-1}$, respectively. The systemic radial 
velocity of Hen\,3-1475 \citep[$+$40~km\,s$^{-1}$,][]{bh01} is 
within the velocity errors.  Although the outermost component D, 
as well as SE1 in the UV, seems to align with its optical 
counterpart along the spectral direction, we cannot rule out the 
possibility that D also has a high approaching speed, given its 
extension and the low spectral resolution of G230L.  This sets an 
upper limit of $\lesssim$310~km\,s$^{-1}$ for the approaching speed
of D.  Adopting this radial speed and the sky-projected velocity of
D ($\sim$280~km\,s$^{-1}$, see Section~\ref{sec4:1}), we deduce a 
jet inclination ($i$, with respect to the line of sight) 
$\gtrsim$42\degr, which is consistent with the estimate of 
\citet[][$i$=40\degr]{bh01}.  Adopting $i$=40\degr, we corrected 
the velocities of A, B, C, and D to be $\sim$2000, 1600, 1400, and 
$\lesssim$400 km\,s$^{-1}$, respectively. 

Both the [N\,{\sc ii}] and H$\alpha$ emission lines in the 
\emph{HST} STIS G750M optical spectroscopy (in 1999; \citealt{bh01})
show an increase in radial velocity from the base region of the NW 
conical shock, reaching a maximum blueshift of $-$910~km\,s$^{-1}$ 
at the cone tip (i.e., component A), and followed by a small drop 
(by $\sim$110~km\,s$^{-1}$) near the cone tip, and continuous 
decrease in velocity through C to D.  This small drop in velocity 
near the cone tip is reflected by the velocity difference between 
A and B in the UV.  Our radial velocities derived for the UV 
components of NW1, although higher than those found in the STIS 
G750M spectrum, generally follow the same trend (i.e., decreasing 
from the inside out), indicating a sequence of negative speed 
changes within NW1.  The jet might have been accelerated since 1999 
possibly because the central star has become hotter. 

If the NW1 components are moving clumps along the jet, the 
sky-projected velocities of A, B, C, and D should be $\sim$1300, 
1000, 900, and $\lesssim$260 km\,s$^{-1}$ (adopting the above 
radial velocities and $i$=40$\degr$), corresponding to proper 
motions of $\sim$55, 42, 38 and $\lesssim$12~mas\,yr$^{-1}$, 
respectively.  However, A, B, and C are actually unchanged in 
positions, whereas D and SE1 show small radial displacements from 
their positions in the 2009 H$\alpha$ image (Figure~\ref{fig3}b; 
see Section~\ref{sec4:1}).  Such a spatio-kinematic pattern of NW1 
seems to resemble quasi-stationary shocks:  A, B, C, and D may 
actually be shock interfaces where the outflowing gas moves through
with high speeds, and have no (or very small) measurable proper 
motions; each of these clumpy shocks may brighten or fade in time 
owing to variations in density or speed of the gas flow.  If A, B, 
C, and D are indeed the locations of shocks, the gas flow should 
be slowed down at each of these shocks as some kinetic energy is 
converted into heat that locally excites line emission (B.\ Balick 
2018, private communication).  The negative radial velocity gradient
in NW1 may thus be explained as a series of speed losses of the 
(possibly unstable) flow as it propagates outward along the jet. 
This interpretation, although plausible, is still speculative and 
needs careful assessment with comprehensive hydrodynamic 
simulations. 


The outer pair of knots (NW2/SE2 and NW3/SE3) in Hen\,3-1475 was 
previously proposed to be fast-moving ``bullets''; but their almost
tangential motion along the $S$-shaped arms 
(Figure~\ref{fig1}~bottom; see also Section~\ref{sec2}) suggests 
that this is probably not the case.  One sensible interpretation is
that these outer knots are actually bright spots along the curved 
arms, which may be faint lobe walls projected on the sky; these 
bright spots are produced due to the Kelvin--Helmholtz instabilities
as excited by the fast streamlines flowing over the surface of lobe 
walls. 

We analyzed the archival \emph{HST} STIS G750L spectrum and found 
that the optical line ratios (e.g., [O\,{\sc ii}]/H$\beta$, [O\,{\sc
iii}]/H$\beta$, [S\,{\sc ii}]/H$\alpha$) from the NW1 knot are 
roughly consistent with the predicted values of a bow-shock 
model at 200~km\,s$^{-1}$ \citep{hartigan87}; this analysis will be 
presented separately (X.\ Fang et al.\ 2018, in preparation). 
The high-velocity jet impinges on the slower AGB wind, creating 
shocks as delineated by the [N\,{\sc ii}]/H$\alpha$ ratio 
(Figure~\ref{fig4}e): the three pairs of knots, as well as the two 
limb-brightened inner cones, are enhanced in [N\,{\sc ii}] emission 
(although we are aware that at least for SE1, the F658N filter may 
be contaminated by redshifted H$\alpha$). 
Within the NW lobe, there are filaments enhanced in the 
H$\alpha$/H$\beta$ ratio (Figure~\ref{fig4}f); they may not be 
due to higher local extinction but actually related to variation 
in shock speed that causes locally enhanced temperature and/or 
density, which results in departure from the Case~B recombination 
of hydrogen.  The NW conical shock is slightly curved before it 
converges (Figure~\ref{fig4}b,\,c), an effect possibly due to 
interaction with the lobe walls, which are bright in scattered 
lights as seen in F555W (Figure~\ref{fig4}d).  The SE conical 
shock is also curved in optical line emission. 

For each of the four striations within the NW lobe, its 1D NUV 
spectrum is composed of a smooth continuum overlaid by several 
emission features, one of them being Mg\,{\sc ii} $\lambda$2800 
(see Figure~\ref{fig4}g for 1D spectra of the inner three, namely 
Striations 1, 2 and 3).  The smooth continuum could be the 
scattered starlight, while the broad UV emission features may come 
from fast stellar outflows.  The innermost UV striation is not 
located exactly on the central star, but offset by 0\farcs13 
(probably due to obscuration by the central dusty torus), 
corresponding to $\sim$9.7$\times$10$^{15}$~cm; the half-width 
(FWHM/2) of Mg\,{\sc ii} emission is $\sim$22\,{\AA}, corresponding
to a radial velocity gradient up to 2360~km\,s$^{-1}$.  These two 
quantities are consistent with both location ($\sim$10$^{16}$~cm) 
and velocity ($\sim$2300~km\,s$^{-1}$) of the ultra-fast 
``pristine'' post-AGB outflow found by \citet{sanchez01}. We cannot
rule out the possibility that such UV spread/broadening might also 
be (at least partially) due to spatial extent; but the broadening 
of emission features on the innermost layer (Striation~1) can
be due to kinematics because its Mg\,{\sc ii} emission seems to 
extend beyond the $U$-shaped boundary of the NW lobe 
(Figure~\ref{fig4}a). 

\subsection{Physical Conditions within the NW1 Knots}
\label{sec4:3}

The exact location of X-ray emission in NW1 is still unclear; 
although it seems to peak around the cone tip (Figure~\ref{fig2}a),
this might be an effect of heavy smoothing of the low-photon 
counting \emph{Chandra} data.  \emph{If} X-ray emission comes 
from the component A, the simultaneous presence of X-ray, UV, and 
optical emission suggests that A is a cooling region of shocked gas
with very strong temperature gradients.  However, A is spatially 
unresolved in our UV monochromatic imaging, given that its angular 
size is comparable to the angular resolution of the STIS/NUV-MAMA 
detector ($\sim$0\farcs06; Table\,\ref{obs_log}).  The physical 
conditions within the NW1 knots will be assessed by a 
re-analysis of the X-ray data and substantiated with the UV and 
optical emission line ratios to derive shock properties (J.\ A.\ 
Toal\'{a} et al.\ 2018, in preparation). 

Using equation~1 of \citet[][]{gruendl04} and the observed Mg\,{\sc 
ii} $\lambda$2800 line flux (5.2$\times$10$^{-14}$ 
erg\,cm$^{-2}$\,s$^{-1}$), we estimate a density of 
$\sim$10$^{5}$~cm$^{-3}$ for component A.  Here an extinction of 
$A_{\rm V}\sim$3.7~magnitude \citep{riera95} and the solar 
abundance \citep[Mg/H=3.98$\times$10$^{-5}$,][]{asplund09} were 
adopted; we also assume a maximum ionization fraction 
Mg$^{+}$/Mg=0.69 at $T_{\rm e}\approx$15\,800~K in ionization 
equilibrium \citep{ss82}, and a filling factor of 1.  Using the 
same assumptions above, we estimate densities of 
$\sim$50\,000--60\,000~cm$^{-3}$ in B, C, and D. 
These estimates of the density from the UV-emitting region in NW1 
are much higher than the electron density derived from the 
[S\,{\sc ii}] $\lambda$6716/$\lambda$6731 ratio 
\citep[$\sim$3000~cm$^{-3}$,][]{riera06}, suggesting 
different physical conditions in the UV, optical, and X-ray 
emission regions (especially in component A). 





\section{Summary and Future Work}
\label{sec5}

We report on monochromatic imaging of Hen\,3-1475 in UV nebular 
emission lines obtained through the \emph{HST} STIS UV long-slit 
spectroscopy; this is the first of such attempt ever made for a 
pPN.  We have analyzed the UV morphology of Hen\,3-1475, in 
conjunction with archival \emph{HST} optical data.  The high spatial
resolution of STIS UV imaging resolves the NW1 knot near the tip of 
the inner conical shock in Hen\,3-1475 into four components in the 
Mg\,{\sc ii} $\lambda$2800 spectral-line image.  These four UV 
components are distributed roughly along the jet axis with slight 
curvature; similar configuration also seems to exist near the SE 
cone.  Compared to their optical counterparts, the four components 
of the NW1 knot are mostly blueshifted due to their high 
(approaching) radial velocities, from $-$1550~km\,s$^{-1}$ on the 
innermost component to $\sim-$300~km\,s$^{-1}$ on the outermost 
one.  Despite of their high radial velocities, the four components 
of NW1 show no obvious proper motions through the period of 
two-decade \emph{HST} observations, indicating that they might be 
quasi-stationary shocks where fast gas flows through.  Given 
this interpretation, the negative radial velocity gradient found in 
NW1 might be explained as successive speed losses of the fast jet 
as it propagates outward, losing kinetic energy at the shock sites 
that are now identified as the four UV components of NW1. 
Moreover, through proper motion studies we found that the outer 
pairs of knots (NW2/SE2 and NW3/SE3) of Hen\,3-1475 are moving 
almost tangentially along the $S$-shaped arms, which indicates 
that they may not be fast-moving ``bullets'' as previously 
suggested, but bright spots induced by Kelvin--Helmholtz 
instabilities. 

The extremely fast conical outflow with the knotty structure near 
its vertex makes Hen\,3-1475 unique from all other pPNe and young 
PNe so far observed.  This Letter is mostly descriptive.  Follow-up
studies of this object, in particular jet collimation and 
interaction with the ambient outflowing AGB gas, based on 
multi-wavelength (X-ray, UV, optical, infrared) observations and 
utilizing state-of-the-art (magneto)hydrodynamic simulations, will 
be carried out (X.\ Fang et al.\ 2018, in preparation).  Given that
Hen\,3-1475 is in rapid evolution, which causes drastic change in 
morphology and physical conditions, close monitoring of this pPN 
with high spatial resolution is essential in order to better 
understand its jet formation/collimation and more interestingly, 
nature of the central star (possibly a binary) and the 
circumstellar/circumbinary environment, such as dust, the disk, 
and the magnetic field.

\acknowledgments

We thank the referee, Jacco van Loon, for very constructive 
suggestions that greatly improved this article.  We thank 
Mart\'{i}n A.\ Guerrero and Bruce Balick for in-depth discussion 
and valuable suggestions during paper drafting and revision.  We 
also thank Albert A.\ Zijlstra and Vincent Icke for comments and 
suggestions.  
J.A.T.\ is funded by UNAM DGAPA PAPITT project IA100318. 
A.I.G.d.C.\ acknowledges support from the Spanish Ministry of 
Economy and Competitiveness through grants ESP2014-54243-R and 
ESP2015-68908-R.  X.F.\ acknowledges the support and hospitality 
of the IAA-CSIC for an academic visit in 2017 September, during 
which this article was discussed and drafted. 
\vspace{1mm}

{\it Facility:} \emph{HST} (STIS) 

%

\end{document}